\documentstyle[12pt,epsfig]{article}

\begin{document}
\begin {center}
{\Large \bf {  Effect of in-medium parameters of $\rho$ meson
              in its photoproduction reactions on nuclei} }
\end {center}
\begin {center}
Swapan Das \footnote {Electronic address: swapand@barc.gov.in} \\
{\it Nuclear Physics Division,
Bhabha Atomic Research Centre  \\
Mumbai-400085, India }
\end {center}

\begin {abstract}
There exist model calculations showing the modification of the hadronic
parameters of $\rho$ meson in the nuclear environment. From these
parameters, we extract the $\rho$ meson nucleus optical potential and
show the medium effect due to this potential on the $\rho$ meson mass
distribution spectra in the photonuclear reaction.
The
calculated results reproduced reasonably the measured $e^+e^-$ invariant
mass, i.e., $\rho$ meson mass, distribution spectra in the
$( \gamma, \rho^0 \to e^+e^- )$ reaction on nuclei.
\end {abstract}


PACS number(s): 25.20.Lj, 13.60.Le

\medskip

\section {Introduction}

The medium modification of the vector meson is a topic of intense interest
in nuclear physics \cite{redv}.
First
indication of the $\rho$ meson modification was seen in the
ultra-relativistic heavy ion collision data taken at CERN-SPS \cite{dilpex}.
Recent
past, STAR experiment at RHIC-BNL \cite{rhic} found that the decrease
in the mass of $\rho$ meson emitted from Au+Au collision is $\sim 70$ MeV.
Contrast
to this, the upgraded CERES experiment \cite{uceres} and the dimuon
measurements in NA60 experiment at CERN \cite{na60} reported
significant broadening (without mass-shift) in the $\rho$ meson mass
distribution spectrum.

It is advantageous to search the in-medium properties of vector meson
in the normal nucleus over those in the ultra-relativistic
heavy-ion collision, since the interpretations of results can be made
more judiciously in the previous case where as those in the latter
case are very difficult.
The
model calculations, e.g.,
scaling hypothesis due to Brown and Rho \cite{br},
QCD sum rule due to Hatsuda and Lee \cite{hat},
vector meson dominance due to Asakawa et al. \cite{aks},
 ..... etc,
envisage the reduction of vector meson mass in the nucleus.
Boreskov
et al., \cite{bore} described the broadening of hadronic resonance in
the nuclear medium. Using intranuclear cascade (INC) approach, Golubeva
et al., \cite{golu} studied the medium modification of vector meson in
the pion nucleus reaction.
In
another calculation, Kundu et al., \cite{kund} showed strong medium
effect on the $\rho$ meson in heavy-ion scattering in GeV region.
The
nuclear medium effect on the vector meson is predicted to be large enough
to be observed it in the photon and hadron induced nuclear reactions
\cite{effn, wdmn, ownc}.

Experimentally too, the medium modification of vector meson produced
in the lepton and hadron induced nuclear reactions had been reported
from various laboratories.
The
$e^+e^-$ yield in $p+$A reaction at 12 GeV (measured by KEK-PS E325
collaboration at KEK \cite{kek}) is reproduced by the mass reduction of
vector meson in the nucleus.
The
$\rho$ meson production experiment in the $( \gamma, \pi^+\pi^- )$
reaction on the nucleus, done by TAGX collaboration \cite{lohu}, reported
the modification of this meson produced in the nucleus.
The
measurements on the nuclear transparency ratio \cite{tpcy} showed large
in-medium width of the $\omega$ and $\phi$ mesons. The in-medium
renormalization (i.e., collision broadening) of the $\omega$ meson was also
seen in the near-threshold  photoproduction of this meson \cite{thiel}.
No
mass-shift but significant broadening of the $\rho$ meson in the
photonuclear reaction has been reported by CLAS collaboration \cite{tjnaf}.
This meson was probed by its decay product $e^+e^-$ in the momentum
range 0.8 to 3.0 GeV/c, and it was produced in the nucleus by
$0.61-3.82$ GeV tagged photon beam at Jefferson Laboratory.
The
experimental results on this topic, as reported by various collaborations,
are tabulated in Ref.~\cite{metag4}.

We consider $e^+e^-$ emission in the photonuclear reaction (quoted
above \cite{tjnaf}) arising due to the elementary reaction in the nucleus:
$ \gamma N \to \rho^0 N ; ~ \rho^0 \to e^+e^- $.
The
$\rho$ meson photoproduction is described by the measured reaction
amplitude $ f_{\gamma N \to \rho N} $. The modification of this meson
occurs due to its interaction with the nucleus.
In
Ref.~\cite{das2}, we developed this interaction or potential using
$``t\varrho"$ approximation, and used it to calculate the $\rho$ meson
mass, i.e., $e^+e^-$ invariant mass, distribution spectra in the
$( \gamma, \rho^0 \to e^+e^- )$ reaction on nuclei.
The
$\rho$ meson potential can also be extracted from its mass-shift and
collision broadening in the nucleus. Using the latter potential, we
calculate the $\rho$ meson mass distribution spectra in the above
reaction and present them along with the data reported from Jefferson
Laboratory \cite{tjnaf}.

\section {Formalism}

The differential cross section of the $\rho$ meson mass $m$ distribution
in the $( \gamma, \rho^0 \to e^+e^- )$ reaction on a nucleus \cite{das2}
can be written as
\begin{equation}
\frac{ d\sigma (E_\gamma) }{ dm }
= \int d\Omega_\rho K_F \Gamma_{\rho^0 \to e^+ e^-} (m)
                   | F({\bf k}_\gamma, {\bf k}_\rho) |^2,
\label{dsc1}
\end{equation}
where $K_F$ denotes the kinematical factor of the reaction:
$ K_F = \frac{3\pi}{(2\pi)^4} \frac{ k^2_\rho E_{A^\prime} m^2 }
  { E_\gamma | k_\rho E_i - {\bf k}_\gamma \cdot {\hat k}_\rho E_\rho | } $;
with $ {\bf k}_\rho = {\bf k}_{e^+} + {\bf k}_{e^-} $.
$E_i$
is the total energy in the initial state. $E_\rho$ and $E_{A^\prime}$
are the energy of the $\rho$ meson and recoil nucleus respectively.
$ \Gamma_{\rho^0 \to e^+e^-} (m) $ stands for the width of the
$\rho$ meson of mass $m$ decaying at rest into dilepton:
$ \Gamma_{\rho^0 \to e^+e^-} (m) \simeq 8.8 \times 10^{-6} m $
\cite{rwde}.
$ F ({\bf k}_\gamma, {\bf k}_\rho) $ describes the photoproduction and
propagation of the $\rho$ meson in the nucleus, i.e.,
\begin{equation}
F ({\bf k}_\gamma, {\bf k}_\rho)
= \int d{\bf r} \Pi_{\gamma A \to \rho A} ({\bf r})
   e^{ i({\bf k}_\gamma - {\bf k}_\rho) . {\bf r} }
  \int^\infty_z dz^\prime D_{\bf k_\rho} (m; {\bf b}, z^\prime, z).
\label{fdfn}
\end{equation}
$ D_{\bf k_\rho} (m; {\bf b}, z^\prime, z) $ in this equation carries
the information about the in-medium properties of $\rho$ meson as it
appears in the $\rho$ meson propagator, see below.

The propagation of the $\rho$ meson from its production point ${\bf r}$
to its decay point ${\bf r^\prime}$ can be expressed as
$ ( -g^\mu_{\mu^\prime} + \frac{1}{m^2} k^\mu_\rho k_{\rho,\mu^\prime} )
G_\rho ( m; {\bf r^\prime - r} ) $ \cite{das2}.
Since
the kinetic energy of this meson (of momentum 0.8 to 3.0 GeV/c) is much
larger than its potential energy, this meson emits dominantly in the
forward direction \cite{das2}.
We,
therefore, represent the scalar part of the in-medium $\rho$ meson
propagator $ G_\rho ( m; {\bf r^\prime - r} ) $ by the eikonal form,
i.e.,
$ G_\rho ( m; {\bf r^\prime - r} ) = \delta ( {\bf b^\prime - b} )
\Theta (z^\prime - z) e^{ i {\bf k_\rho \cdot (r^\prime - r)} }
D_{\bf k_\rho} ( m; {\bf b}, z^\prime, z ) $ \cite{das2},
where
$ D_{\bf k_\rho} ( m; {\bf b}, z^\prime, z ) $ is given by
\begin{equation}
D_{\bf k_\rho} ( m; {\bf b}, z^\prime, z ) =
-\frac{i}{ 2k_{\rho\parallel} }
exp \left [  \frac{i}{ 2k_{\rho\parallel} } \int_z^{z^\prime}
 dz^{\prime \prime} \{ {\tilde G}^{-1}_{0\rho} ( m ) -
    2 E_\rho V_{O\rho} ({\bf b}, z^{\prime \prime}) \} \right  ].
\label{dom}
\end{equation}
$ V_{O\rho} ({\bf b}, z^{\prime \prime}) $ is the $\rho$ meson nucleus
optical potential which can modify the hadronic parameters of this
meson during its passage through the nucleus.
$ {\tilde G}_{0\rho} (m) $ denotes the free space $\rho$ meson 
propagator: $ {\tilde G}^{-1}_{0\rho} (m)
= m^2-m^2_\rho+im_\rho\Gamma_\rho (m) $, where $m_\rho$ is the pole
mass of this meson.
In
fact, the above form of the in-medium propagator was used to study the
medium effects on the vector meson produced in the nuclear reactions
\cite{golu, kund}.

$ \Pi_{\gamma A \to \rho A} ( {\bf r} ) $ in Eq.~(\ref{fdfn}) represents
the generalized optical potential or self-energy of the $\rho$ meson which
describes its production in the photonuclear reaction
\cite{das2, pash}:
\begin{equation}
\Pi_{\gamma A \to \rho A} ( {\bf r} ) = -4\pi E_\rho
\left [ \frac{1}{{\tilde E}_\rho} + \frac{1}{\tilde {E}_N} \right ]
\tilde {f}_{\gamma N \to \rho N} (0) \varrho ({\bf r}),
\label{gpa}
\end{equation}
where $\varrho ({\bf r})$ denotes the spatial density distribution
of the nucleus. This distribution is described by two parameter Fermi
distribution function for all considered nuclei except $^{12}$C, for
which it is represented by the harmonic oscillator gaussian distribution
\cite{das2, andt}.
$ \tilde {f}_{\gamma N \to \rho N} (0) $ is the amplitude of the
$\gamma N \to \rho N$ reaction. Its magnitude square, which appears in
Eq.~(\ref{dsc1}), can be extracted directly from the measured
differential cross section \cite{sath}:
$ \frac{d\sigma}{dt} (\gamma N \to \rho N)|_{t=0}
\simeq \frac{\pi}{{\tilde k}^2_\gamma}
| \tilde {f}_{\gamma N \to \rho N} |^2 $.
The symbol tilde on the quantities represents those in the $\rho N$ c.m.
system of energy equal to that in the $\gamma N$ c.m. system.

Eq.~(\ref{dsc1}) can be used to estimate the differential cross section
of the $ \rho^0 (\to e^+e^-) $ meson mass distribution due to fixed
$\gamma$ beam energy $E_\gamma$. But the $\rho$ meson, as mentioned
earlier, was produced by the tagged photon beam \cite{tjnaf} whose
profile (illustrated in Ref.~\cite{rrol}) can be divided into 6 energy
bins. Therefore, the cross section can be expressed as
$ \frac{d\sigma}{dm} = \sum^6_{i=1}
W(E_{\gamma,i}) \frac{d\sigma (E_{{\gamma,i}})} {dm} $,
where $d\sigma(E_{\gamma,i}) / dm$ is given by Eq.~(\ref{dsc1}).
$E_{\gamma,i}$ consists of six bins of photon energies,
$E_{\gamma,i}$ (GeV) = 1.0, 1.5, 2.0, 2.5, 3.0 and 3.5 with relative
weights $W(E_{\gamma,i})$ of $13.7\%$, $23.5\%$, $19.3\%$, $20.1\%$,
$12.6\%$ and $10.9\%$ respectively \cite{rrol}.

\section {Results and Discussion}

The optical potential of the $\rho$ meson can be extracted from its
modified mass $ \Delta m $ and width $ \Delta \Gamma $ in the nucleus
\cite{effn}. 
The
latter, i.e., $\Delta \Gamma$, due to its mass modification can be
neglected in compare to its collision broadening \cite{wdmn}. The
$\rho$ meson optical potential is related to its mass and width
\cite{kund, elet} as
\begin{equation}
\Delta m ({\bf r}) = m^* ({\bf r}) - m  
= \gamma_L \frac{E_\rho}{m} Re V_{O\rho} ({\bf r});
~~~~~
\Delta \Gamma ({\bf r}) = \Gamma^* ({\bf r}) - \Gamma  
= - \gamma_L \frac{2E_\rho}{m} Im V_{O\rho} ({\bf r}),
\label{cmwad}
\end{equation}
where $\gamma_L$ is the associated Lorentz factor \cite{kscge}. $m$
and $\Gamma$ are the mass and width of the $\rho$ meson in free space.
The asterisk stands for the in-medium quantities. In fact, these
relations are described for the pole mass of the $\rho$ meson. We assume
they do not vary with the mass of this meson.

The QCD sum rule calculation due to Hatsuda and Lee \cite{hat} have
predicted the linear variation of $ m^* ({\bf r}) $ with the nuclear
density $ \varrho ({\bf r}) $. According to them, it is 
\begin{equation}
m^* ({\bf r}) = m [ 1 - 0.18 \varrho({\bf r}) / \varrho(0) ].
\label{mhtle}
\end{equation}
On the other hand,
Asakawa et al., \cite{aks} in their vector meson dominance (VMD) model
calculation have shown that $ m^* ({\bf r}) $ varies non-linearly with
$ \varrho ({\bf r}) $, i.e.,
\begin{equation}
m^* ({\bf r}) = m / [ 1 + 0.25 \varrho({\bf r}) / \varrho(0) ].
\label{maskw}
\end{equation}

The in-medium mass $m^*$ of the $\rho$ meson, as shown by Kondratuyk
et al. \cite{kscge}, increases with the logarithmic increase in its
momentum $k_\rho$ such that
(i) $ m^* \sim 0.82 m $ at the central density of the nucleus as
$ k_\rho \to 1 $ MeV (see Eq.~(\ref{mhtle})),
and
(ii) the mass-shift $ \Delta m $ of this meson is zero at
$ k_\rho \approx 100 $ MeV/c.
Effenberger et al., \cite{effn} has incorporated such behavior by
multiplying the function
$ f(k_\rho) = \left ( 1 - \frac{ k_\rho }{ 1 ~\mbox{GeV/c} } \right ) $
with the mass-shift (or the real part of optical potential) of the $\rho$
meson.
Since
this form is linear in $k_\rho$, it cannot reproduce $k_\rho$
dependence of $m^*$ as described by Kondratuyk et al., \cite{kscge}.
In addition, this form of $f(k_\rho)$ distinctly fails to satisfy the
second condition stated above. 
To
overcome these shortcomings, we propose a new form of $f(k_\rho)$ which
duly satisfies the behavior of $m^*$ as elucidated above in (i) and (ii):
\begin{equation}
f ( k_\rho )
= - \frac{ 1 }{ 2 } \left ( 1 + log \frac{k_\rho}{k^0_\rho} \right );
~~~~~ k^0_\rho = 1 ~\mbox{GeV/c}.
\label{fkp}
\end{equation}

The parametric form of the density dependent $\rho$ meson width
$\Gamma^*_\rho ({\bf r})$, which is well accord with its dynamically
generated width, is given by Weidmann et al. \cite{wdmn} only for
$ k_\rho \leq 1 $ GeV/c. Nothing about it is interpreted beyond this
momentum. According to them, $ \Gamma^*_\rho ({\bf r}) $ can be
expressed as
\begin{equation}
\Gamma^*_\rho ({\bf r})
= \Gamma_\rho [ 1 + 1.55 \varrho ({\bf r}) / \varrho (0) ].
\label{wptr}
\end{equation}
Based
on CLAS experimental results reported for $ k_\rho = 0.8-3.0 $ GeV/c,
the nuclear density dependence of $ \Gamma^*_\rho ({\bf r}) $
is illustrated by Leupold et al. \cite{redv} as
\begin{equation}
\Gamma^*_\rho ({\bf r})
= \Gamma_\rho [ 1 + \varrho ({\bf r}) / \varrho (0) ].
\label{wpcl}
\end{equation}
The in-medium width $ \Gamma^*_\rho ({\bf r}) $, as worked out in
Ref.~\cite{kscge}, weakly varies with the $\rho$ meson momentum.

We extract the $\rho$ meson optical potential from its modified mass
and width elucidated in Eqs.~(\ref{cmwad}) - (\ref{wpcl}), and use
them to calculate the $\rho$ meson mass distribution spectra in the
$( \gamma, \rho^0 \to e^+ e^- )$ reaction on nuclei for
$ k_\rho = 0.8 - 3.0 $ GeV/c.
The
calculated spectra due to $ Re V_{O\rho} ({\bf r}) $ evaluated from
$m^*$ in Eq.~(\ref{mhtle}) do not differ much from those due to
$ Re V_{O\rho} ({\bf r}) $ estimated from $m^*$ in Eq.~(\ref{maskw}).
Hence,
we do not present the latter results.

\begin{figure}[h]
\begin{center}
\centerline {\vbox {
\psfig{figure=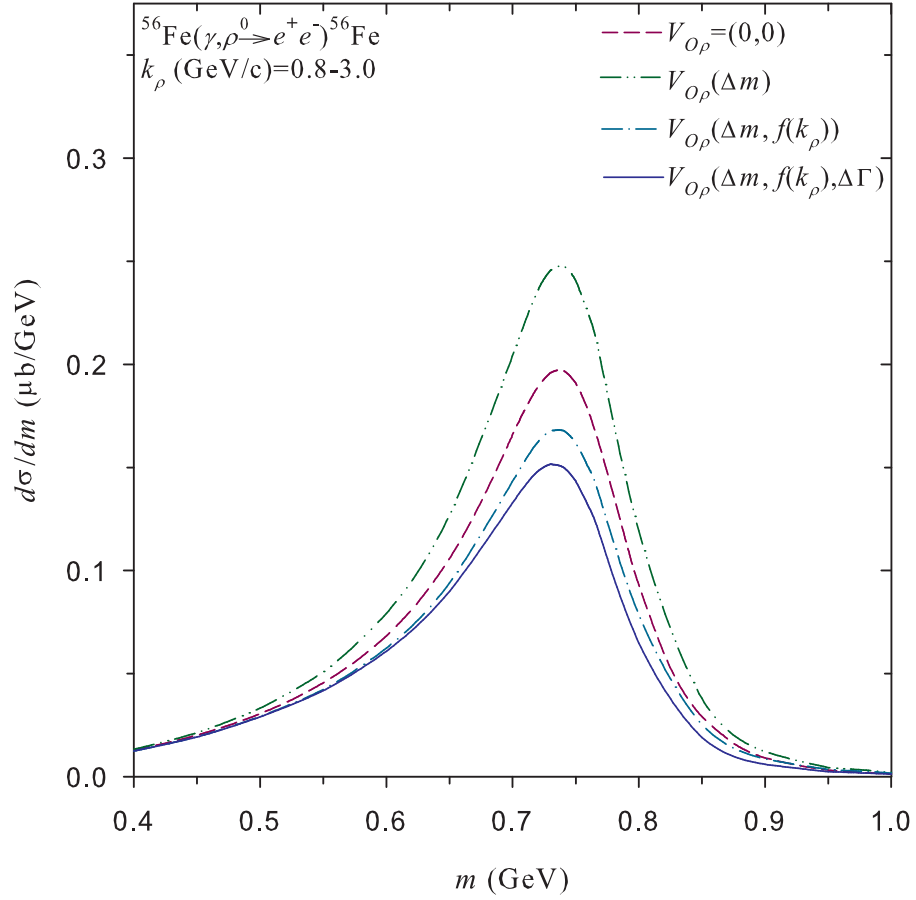,height=12.0 cm,width=12.0 cm}
}}
\caption{
(color online).
The calculated $\rho$ meson mass distribution spectra due to its
potentials extracted from its mass-shift $ \Delta m $ and collision
broadening $ \Delta \Gamma $ (see text).
}
\label{fhlFe}
\end{center}
\end{figure}

The short-dash curve in Fig.~\ref{fhlFe} represents the $\rho$ meson
mass distribution spectrum for its free propagation, i.e.,
$ V_{O\rho} = ( 0,0) $, through $^{56}$Fe nucleus.
This
curve shows that the peak cross section of the considered reaction is
$ \sim 0.2 ~\mu$b/GeV which appears at the $\rho$ meson mass equal to
740 MeV.
The
dot-dot-dash curve refers to above spectrum arising due to the
incorporation of the $\rho$ meson optical potential
$ V_{O\rho} ( \Delta m ) $, extracted from the mass-shift $\Delta m$
only (see Eqs.~(\ref{cmwad}) and (\ref{mhtle})).
It
shows the cross section due to this potential is enhanced by a factor
of $\sim 1.26$ at the peak.
In
this figure, $ V_{O\rho} ( \Delta m, f(k_\rho) ) $ illustrates the
inclusion of the $\rho$ meson momentum $k_\rho$ dependence in
$ V_{O\rho} ( \Delta m ) $ as described in Eq.~(\ref{fkp}).
The
dot-dash curve shows that the peak cross section is reduced by a
factor of $\sim 1.5$ because of this
potential.
$ V_{O\rho} ( \Delta m, f(k_\rho), \Delta \Gamma ) $ represents the
$\rho$ meson optical potential extracted from the momentum dependent
mass-shift and collision broadening (discussed in Eqs.~(\ref{cmwad})
- (\ref{wpcl})).
Because
of this potential, the calculated cross section (shown by
the solid curve) is reduced further by a factor of 1.11 at the peak.
This
figure also shows that the peak cross section due to the latter
potential is reduced by 1.3 in compare to that (short-dash curve) obtained
because of the non-interacting $\rho$ meson propagation through the
nucleus.

\begin{figure}[h]
\begin{center}
\centerline {\vbox {
\psfig{figure=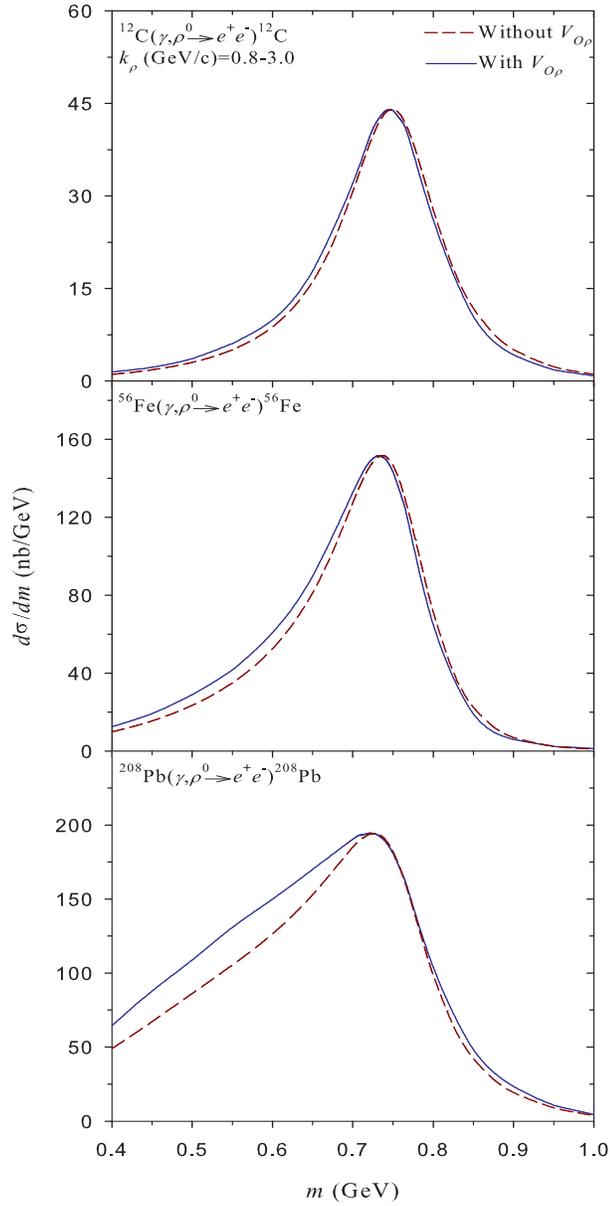,height=16.0 cm,width=8.0 cm}
}}
\caption{
(color online).
Broadening, but essentially negligible peak-shift, in the $\rho$ meson
mass distribution spectra occurring due to its potential $V_{O\rho}$.
The dashed curve is divided by 1.29 for $^{12}$C; 1.3 for $^{56}$Fe
and 1.35 for $^{208}$Pb.
}
\label{fhlmd}
\end{center}
\end{figure}

\begin{figure}[h]
\begin{center}
\centerline {\vbox {
\psfig{figure=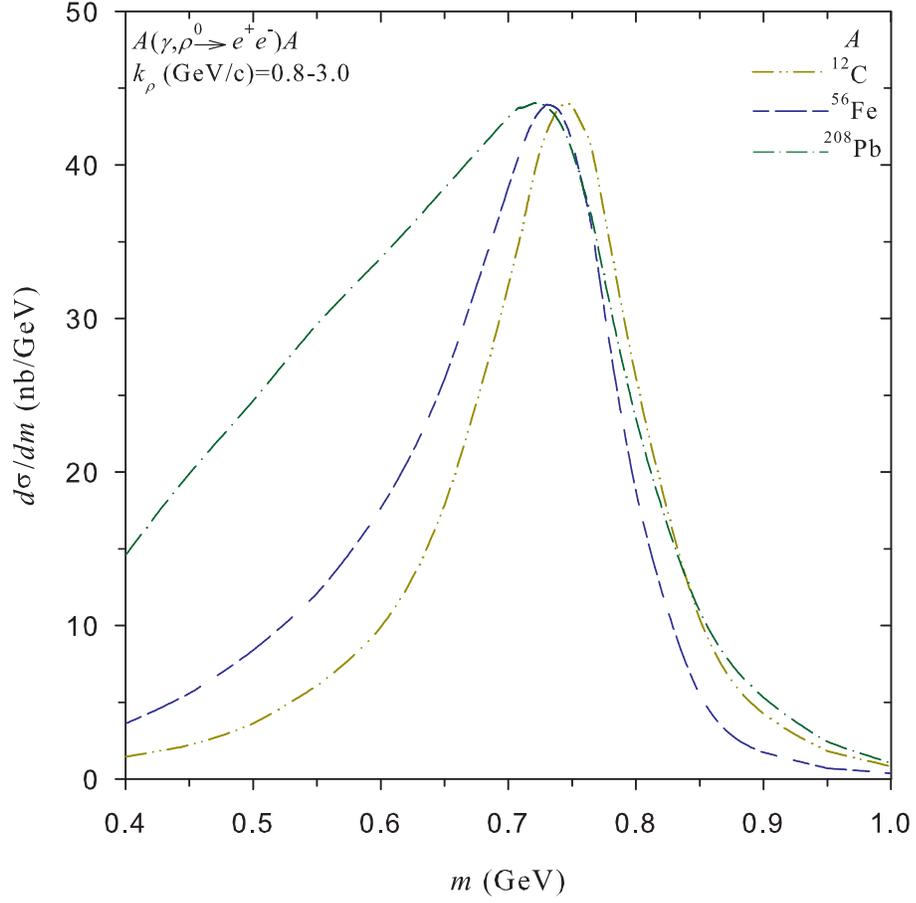,height=12.0 cm,width=12.0 cm}
}}
\caption{
(color online).
Enhancement in the $\rho$ meson width with the size of the nucleus. The
shift of its mass towards the lower value in the heavier nucleus is
insignificant.
The short-long-short and dot-dash curves are divided by the factors 3.46
and 4.44 respectively.
}
\label{fhlNu}
\end{center}
\end{figure}

The modification of the $\rho$ meson parameters in the nuclear reaction
can occur because of its interaction with the nucleus, i.e., $V_{O\rho}$
in Eq.~(\ref{cmwad}).
Therefore,
the calculated results with and without this potential can illustrate
the medium modification of this meson in the considered reaction,
i.e., $( \gamma, \rho^0 \to e^+e^- )$ reaction on nuclei.
The
solid and dashed curves in Fig.~\ref{fhlmd} represent the calculated
$\rho$ meson mass distribution spectra with and without $V_{O\rho}$
respectively.
It
elucidates the broadening in the $\rho$ meson mass distribution spectrum
occurring (more distinctly in heavy nucleus) due to $V_{O\rho}$, without
significant peak-shift (within 10 MeV).
In
Fig.~\ref{fhlNu}, we compare the $\rho$ meson mass distribution spectra
calculated for various nuclei. These spectra show insignificant shift
of the $\rho$ meson peak-mass but remarkable enhancement in its width
with the size of the nucleus.

\begin{figure}[h]
\begin{center}
\centerline {\vbox {
\psfig{figure=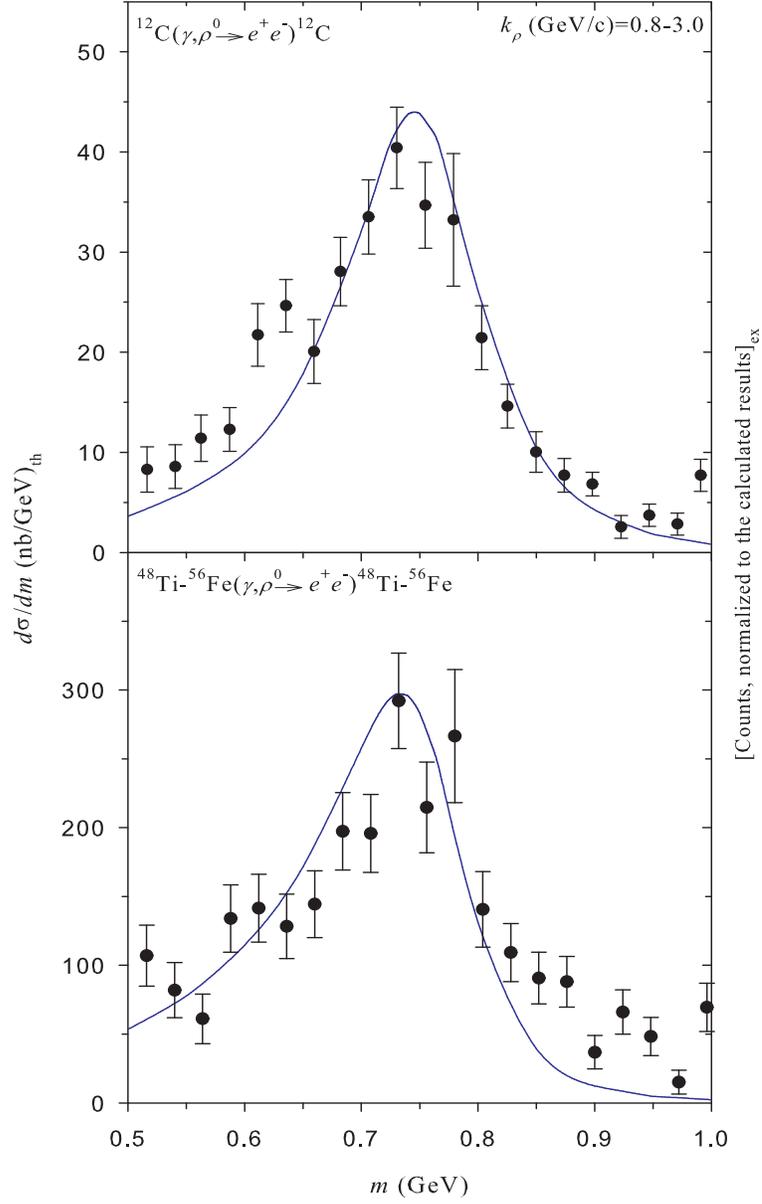,height=16.0 cm,width=10.0 cm}
}}
\caption{
(color online).
The calculated results (solid curve) compared with the data taken from
Ref.~\cite{tjnaf}.
}
\label{fhlDt}
\end{center}
\end{figure}

The calculated $\rho$ meson mass distribution spectrum for $^{12}$C
nucleus along with the data is shown in the upper part of
Fig.~\ref{fhlDt}.
Those
for $^{48}$Ti and $^{56}$Fe nuclei (combined) are compared with the
data in the lower part of this figure. It shows that the calculated
results are accord with the measured distributions reported from
Jefferson laboratory \cite{tjnaf}.

\begin{figure}[h]
\begin{center}
\centerline {\vbox {
\psfig{figure=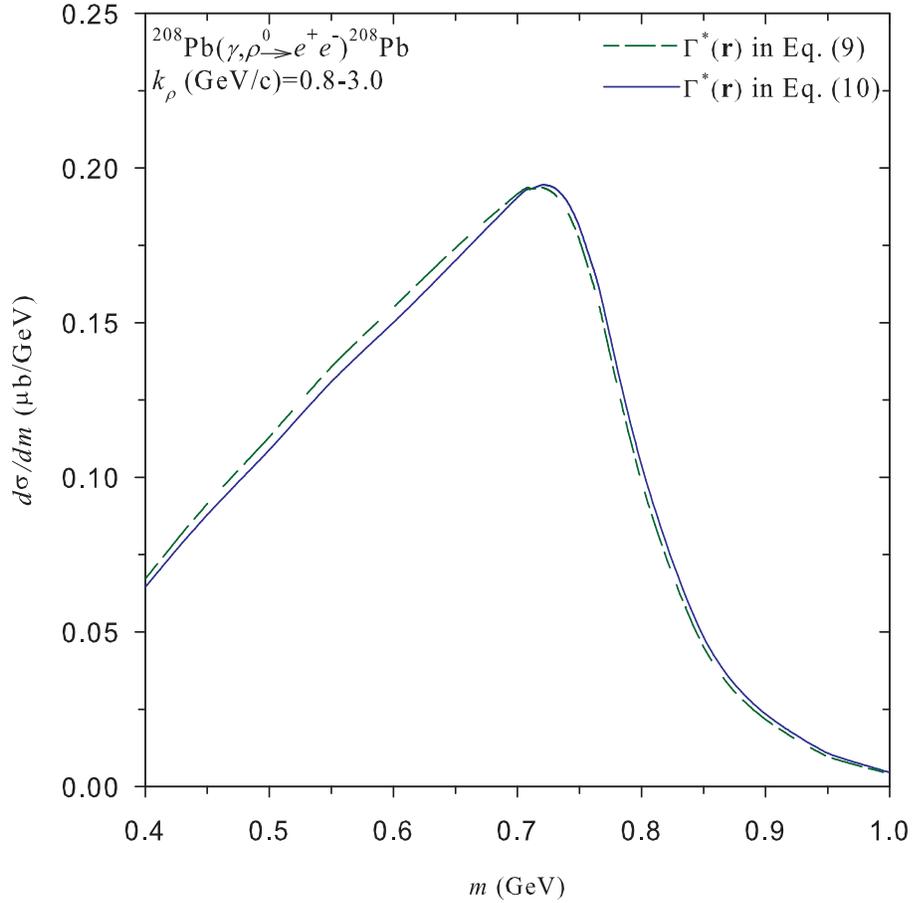,height=12.0 cm,width=12.0 cm}
}}
\caption{
(color online).
The sensitivity of the $\rho$ meson mass distribution spectra to its
in-medium widths given in Eqs.~(\ref{wptr}) and (\ref{wpcl}). The
short-long-short dash curve is obtained because of Eq.~(\ref{wptr}) for
$ \Gamma^* ({\bf r}) $ where as the solid curve arises due to
$ \Gamma^* ({\bf r}) $ given in Eq.~(\ref{wpcl}). The previous is
multiplied by 1.04.
}
\label{fhlww}
\end{center}
\end{figure}

All results, presented so far, are calculated using the imaginary part
of the $\rho$ meson potential $ Im V_{O\rho} ({\bf r}) $ extracted
from its in-medium width $ \Gamma^*_\rho ({\bf r}) $ given in
Eq.~(\ref{wpcl}).
To
disentangle the effect of $ \Gamma^*_\rho ({\bf r}) $ in Eq.~(\ref{wptr})
on the $\rho$ meson mass distribution spectrum in the considered reaction,
we
calculate this spectrum using $ \Gamma^*_\rho ({\bf r}) $ in
Eq.~(\ref{wptr}) for $^{208}$Pb nucleus where the medium effect on this
meson, as shown in Fig.~\ref{fhlmd}, is the largest.
In
Fig.~\ref{fhlww}, this result (short-long-short dash curve) is compared 
with the previous (solid curve) calculated using
$ \Gamma^*_\rho ({\bf r}) $ in Eq.~(\ref{wpcl}).
This
figure shows that the $\rho$ meson mass distribution spectrum is hardly
sensitive to the different forms of $ \Gamma^*_\rho ({\bf r}) $,
given in Eqs.~(\ref{wptr}) and (\ref{wpcl}).

\section {Conclusions}
We have calculated the differential cross section for the $\rho$ meson
mass distribution in the $( \gamma, \rho^0 \to e^+e^- )$ reaction on
nuclei using the potential extracted from the in-medium parameters of
this meson.
The
calculated spectra show that the broadening (specifically in heavy
nucleus) without significant peak-shift occurs because of the $\rho$
meson nucleus interaction. The calculated spectra reproduce the data
reasonably.
These
results corroborate qualitatively those obtained due to the potential
evaluated using the $\rho$ meson nucleon scattering parameters \cite{das2}.
Depending
on nucleus, the cross sections in the present case is about 2.5 to 3.5
times larger than those reported earlier in Ref.~\cite{das2}.
Therefore,
it is very much necessary to obtain the respective absolute normalized
data which can be used to justify the $\rho$ meson potential. Once it is
confirmed,
the in-medium parameters of $\rho$ meson can be worked out unambiguously.

\section {Acknowledgement}
The author thanks the referee for his/her comments which improve the quality
of this work. R. Nasseripour is acknowledged for sending the data for
Fe-Ti nuclei.

\newpage

{\bf Figure Captions}

\begin{enumerate}

\item (color online).
The calculated $\rho$ meson mass distribution spectra due to its
potentials extracted from its mass-shift $ \Delta m $ and collision
broadening $ \Delta \Gamma $ (see text).

\item (color online).
Broadening, but essentially negligible peak-shift, in the $\rho$ meson
mass distribution spectra occurring due to its potential $V_{O\rho}$.
The dashed curve is divided by 1.29 for $^{12}$C; 1.3 for $^{56}$Fe
and 1.35 for $^{208}$Pb.

\item (color online).
Enhancement in the $\rho$ meson width with the size of the nucleus. The
shift of its mass towards the lower value in the heavier nucleus is
insignificant. The short-long-short and dot-dash curves are divided by
the factors 3.46 and 4.44 respectively.

\item (color online).
The calculated results (solid curve) compared with the data taken from
Ref.~\cite{tjnaf}.

\item (color online).
The sensitivity of the $\rho$ meson mass distribution spectra to its
in-medium widths given in Eqs.~(\ref{wptr}) and (\ref{wpcl}). The
short-long-short dash curve is obtained because of Eq.~(\ref{wptr}) for
$ \Gamma^* ({\bf r}) $ where as the solid curve arises due to
$ \Gamma^* ({\bf r}) $ given in Eq.~(\ref{wpcl}). The previous is
multiplied by 1.04.

\end{enumerate}

\newpage

\end{document}